\documentclass[11pt]{article}
\usepackage{amsmath}
\usepackage{amsfonts}
\usepackage[english]{babel}
\usepackage{amssymb}
\usepackage{amsthm}
\usepackage{graphicx}
\usepackage[a4paper]{geometry}
\usepackage{hyperref}
\usepackage{amstext}
\usepackage{caption}
\usepackage{etoolbox}
\usepackage{makeidx}
\usepackage{amsfonts}
\usepackage{authblk} 
\usepackage{sectsty}
\usepackage{amscd}
\usepackage{mathrsfs}
\usepackage{dsfont}
\usepackage{enumitem} 
\usepackage[]{latexsym}
\usepackage{braket}
\usepackage{caption}

\usepackage{xcolor}

\newcommand{\be}{\begin{equation}}
\newcommand{\ee}{\end{equation}}
\newcommand{\bea}{\begin{eqnarray}}
\newcommand{\eea}{\end{eqnarray}}

\newtheorem*{proof*}{Proof}

\title{A  gentle introduction to Schwinger's formulation of Quantum Mechanics: the groupoid picture}
\author[1,2,5]{F. M. Ciaglia}
\author[3,4,6]{A. Ibort}
\author[1,2,7]{G. Marmo}

\affil[1]{\textit{\footnotesize Dipartimento di Fisica ``E. Pancini'', Universit\`a di Napoli Federico II, Complesso Universitario di Monte S. Angelo Edificio 6, via Cintia, 80126 Napoli, Italy.}}

\affil[2]{\textit{\footnotesize INFN-Sezione di Napoli, Complesso Universitario di Monte S. Angelo Edificio 6, via Cintia, 80126 Napoli, Italy.}}

\affil[3]{\textit{\footnotesize ICMAT, Instituto de Ciencias Matem\'{a}ticas (CSIC-UAM-UC3M-UCM).}}

\affil[4]{\textit{\footnotesize Depto. de Matem\'aticas, Univ. Carlos III de Madrid, Avda. de la Universidad 30, 28911 Legan\'es, Madrid, Spain.}}

\affil[5]{\footnotesize e-mail:\texttt{ florio.m.ciaglia@gmail.com}}

\affil[6]{\footnotesize e-mail:\texttt{ albertoi@math.uc3m.es}}

\affil[7]{\footnotesize e-mail:\texttt{ marmo@na.infn.it}}

\date{}

\begin{document}

\maketitle

\abstract{In this short letter we review Schwinger's formulation of Quantum Mechanics and we argue that the mathematical structure behind Schwinger's ``Symbolism of Atomic Measurements'' is that of a groupoid.
In this framework, both the Hilbert space (Schr\"{o}dinger picture) and the $C^{*}$-algebra (Heisenberg picture) of the system turn out to be derived concepts, that is, they arise from the underlying groupoid structure.}

\section{Introduction}

At the dawn of quantum mechanics, two pictures for the description of quantum systems emerged.
One, proposed by W. Heisenberg, was called matrix mechanics and kept the form of classical Hamiltonian equations for noncommutative quantities, the other, proposed by E. Schr\"{o}dinger, was named wave mechanics and was formulated in analogy with wave equations.
In both of them was clearly present what Dirac called the ``principle of analogy''.
It may be worth mentioning that Dirac in his lectures on Quantum Field Theory \cite{dirac-lectures_on_quantum_field_theory}, talking about Basic Quantum Concepts, clearly says: ``what is the precise mathematical nature of the dynamical variables? It is better, for the present, to keep an open mind about these dynamical variables and just call them q-numbers''.
He then goes on by imposing algebraic axioms that look reasonable, postponing the problem of ``identification as physical entities''.

Few years later, searching for a relativistically covariant description of quantum systems, Dirac proposed a Lagrangian approach to the description of quantum systems \cite{dirac-the_lagrangian_in_quantum_mechanics}.
This few pages paper inspired, separately, both Feynman and Schwinger to provide an alternative formulation, or ``picture'', of quantum mechanics by means of a Lagrangian or an ``action principle'' respectively.
It is quite interesting to notice that when addressing the problem of introducing quantum mechanics at an elementary level (text-book level), both Feynman and Schwinger made recourse to the Stern-Gerlach experiment.
Feynman used it in its presentation of quantum mechanics in the third volume of his famous lectures, and Schwinger used it in his formulation of quantum mechanics in terms of a ``Symbolism of Atomic Measurements'' \cite{schwinger-quantum_mechanics_symbolism_of_atomic_measurements}.

Now, the primary mathematical object for the ``Schr\"{o}dinger picture'' was identified by Dirac in the Hilbert space of the system \cite{dirac-principles_of_quantum_mechanics}, and states, observables and evolution, along with a probability function, where identified as objects built out of the Hilbert space of the system.
While, for the Heisenberg picture, the primary mathematical structure was identified to be a $C^{*}$-algebra assigned to each quantum system, and observables, states, evolution and probability function were clearly identified by von Neumann \cite{von_neumann-mathematical_foundations_of_quantum_mechanics}.
On the other hand, it seems to us that the mathematical structure underlying Schwinger's picture of quantum mechanics  has not yet been identified in its bare essentials.
A thorough extensive presentation will be published elsewhere \cite{ciaglia_ibort_marmo-schwinger}.
In this short note, we would like to argue that the proper mathematical structure emerging from the ``Symbolism of Atomic Measurements'' is that of a groupoid and its higher algebraic generalizations.
In some sense, we could argue that in this picture the probabilistic point of view plays a much more prominent role than in the other two pictures attributed, respectively, to Schr\"{o}dinger and Heisenberg.
In this respect, Schwinger's proposal is closer to the poisitivistic thinking of Heisenberg.

\section{Symbolism of Atomic Measurements and groupoids}

In the incipit of his first note \cite{schwinger-the_algebra_of_microscopic_measurement}, Schwinger  writes: ``The classical theory of measurement is implicitly based upon the concept of an
interaction between the system of interest and the measuring apparatus that can
be made arbitrarily small, or at least precisely compensated, so that one can speak
meaningfully of an idealized experiment that disturbs no property of the system.
The classical representation of physical quantities by numbers is the identification
of all properties with the results of such nondisturbing measurements. It is characteristic
of atomic phenomena, however, that the interaction between system and
instrument cannot be indefinitely weakened. Nor can the disturbance produced by
the interaction be compensated precisely since it is only statistically predictable.
Accordingly, a measurement on one property can produce unavoidable changes in the value previously assigned to another property, and it is without meaning to
ascribe numerical values to all the attributes of a microscopic system. 
The mathematical language that is appropriate to the atomic domain is found in the {\itshape symbolic transcription of the laws of microscopic measurement}\footnote{The emphasizing is due to the authors.}''.

Moved by this operational attitude, Schwinger considers idealized physical systems described by means of a maximal set\footnote{In the general picture discussed in  \cite{ciaglia_ibort_marmo-schwinger} we do not require maximality. %as we do not wish to identify outcomes of experiments with ``states''. 
Such an approach is much more flexible and allows notions like Sorkin's ``events'' \cite{frauca_sorkin-how_to_measure_the_quantum_measure} to be considered in the same ground.} $\mathbf{A}$ of compatible observables  presenting only a finite number of possible outcomes of measurements.
These possible outcomes are denoted as $a,\,a',\,...,\,a^{n}$.
At this point, the so-called measurement symbols $M(a,\,a')$ are introduced without any reference to the Hilbert space and without making recourse to any analogy with classical physics. 
In Schwinger's own words \cite{schwinger-the_algebra_of_microscopic_measurement}, the measurement symbol ``$M(a',\,a)$ indicates a selective measurement in which systems are accepted only in the state $a$ and emerge in the state $a'$''.
In particular, the case in which no change occurs is denoted with $M(a)\equiv M(a,\,a)$.  
The next step consists in giving a composition law for measurement symbols:
\be
M(a''',\,a'')M(a',\,a)=\delta(a',\,a'')\,M(a''',\,a)\,,
\ee
where $\delta$ is the Kronecker delta.
This composition of measurement symbols is postulated and it reflects the fact that when successive measurements are considered, the second stage $M(a''',\,a'')$ of the full measurement scheme accepts the output of the first stage $M(a',\,a)$ if and only if $a'=a''$. 
Quite interestingly, reversing the order in the compound measurement scheme, we obtain $M(a',\,a)M(a''',\,a'')=\delta(a,\,a''')\,M(a',\,a'')$ which means that the composition of measurement symbols is noncommutative.

To make contact with the well-known Dirac-Schr\"{o}dinger formalism, let us consider the association of a Hilbert space with every quantum system, and use the bra-ket formalism\footnote{In the abstract groupoid formalism of \cite{ciaglia_ibort_marmo-schwinger} the Hilbert space is provided gratis as it comes from the fundamental representation of the groupoid.}.
In this setting, Schwinger's maximal set of compatible observables translates into the notion of maximal set of mutually commuting self-adjoint operators on the Hilbert space of the system.
In the finite-dimensional case, we may consider a single observable (self-adjoint operator) $\mathbf{A}$ whose  spectrum is necessarily discrete, say $\{a_{1},\,\cdots,\,a_{n}\}$, and we obtain a resolution of the identity in terms of an orthonormal basis of eigenvectors, that is, $\mathbb{I}=\sum_{j}\,|a_{j}\rangle\langle a_{j}|$  with $\langle a_{j}|a_{k}\rangle=\delta_{jk}$.
It is then clear that  the elements $M(a_{j},\,a_{k})=|a_{j}\rangle\langle a_{k}|$ represent Schwinger's measurement symbols.

When endowed with the associative product inherited from linear operators, the set $\Gamma_{A}$ of such symbols form a groupoid, more precisely, a pair-groupoid \cite{ibort_manko_marmo_simoni_stornaiolo_ventriglia-groupoids_and_the_tomographic_picture_of_quantum_mechanics}.
Indeed, for every $M(a_{j},\,a_{k})\in\Gamma_{A}$ we can define the source and target maps $s,t\colon \Gamma_{A}\rightarrow \Gamma_{A}^{0}$, where $\Gamma_{A}^{0}:\left\{M(a_{j},\,a_{j})\in\Gamma_{A}\right\}$, given by $s(M(a_{j},\,a_{k}))=M(a_{k},\,a_{k})$ and $t(M(a_{j},\,a_{k}))=M(a_{j},\,a_{j})$, and it is immediate to check that the (associative) composition law $M(a_{j},\,a_{k})\circ M(a_{l},\,a_{m})$ coming from the operator product between $M(a_{j},\,a_{k})=|a_{j}\rangle\langle a_{k}|$ and $M(a_{l},\,a_{m})=|a_{l}\rangle\langle a_{m}|$ makes sense whenever $s(M(a_{j},\,a_{k}))= t(M(a_{l},\,a_{m}))$ in which case it is $M(a_{j},\,a_{k})\circ M(a_{l},\,a_{m})=M(a_{j},\,a_{m})$.
Furthermore, every $M(a_{j},\,a_{k})$ has a (right and left) inverse $M^{-1}(a_{j},\,a_{k})=M(a_{k},\,a_{j})$.

Note that, the abstract composition law for measurement symbols is postulated by Schwinger, while here it appears arising from operators simply because of our choice of a representation.
In a forthcoming work \cite{ciaglia_ibort_marmo-schwinger}, we will take up the abstract point of view in order to reformulate Schwinger's ideas in terms of a purely abstract groupoidal picture of quantum mechanics in which the Hilbert space and the $C^{*}$-algebra of (bounded) observables will be ``derived concepts''.

The full algebra of quantum measurements introduced by Schwinger is nothing but the  ``groupoid-algebra'' associated with $\Gamma_{A}$, which is the analogue of the well-known ``group-algebra'' in group theory with the corresponding ``convolution product'' \cite{balachandran_jo_marmo-group_theory_and_hopf_algebras-lectures_for_physicists}.
In the axiomatics presented by Schwinger the ``groupoid'' and its ``groupoid algebra'' are not distinguished, they are introduced at the same time.
The ``objects'' are identified with the unities and the Hilbert space arises from the fundamental representation.
In the bra-ket notation, the groupoid-algebra is given by complex linear combinations of $|a_{j}\rangle\langle a_{k}|$.
An involution $\dagger$ on the groupoid-algebra is immediately defined starting from $M^{\dagger}(a_{j},\,a_{k})=\left(|a_{j}\rangle\langle a_{k}|\right)^{\dagger}=|a_{k}\rangle\langle a_{j}|=M(a_{k},\,a_{j})$ and extending it by linearity.
Real elements will be generated by $\left(|a_{j}\rangle\langle a_{k}| + |a_{k}\rangle\langle a_{j}|\right)$ and $\imath\left(|a_{j}\rangle\langle a_{k}| - |a_{k}\rangle\langle a_{j}|\right)$ with real coefficients, and we already know that there is a unit element $\sum_{j}\,|a_{j}\rangle\langle a_{j}|=\mathbb{I}$.
When $\mathcal{H}$ is finite-dimensional, it is clear that we recover the full $C^{*}$-algebra of bounded linear operators on $\mathcal{H}$.
In the infinite-dimensional case some care must be taken in order to overcome convergence issues of infinite sums.

Schwinger's immediately recognized that ``the physical quantities contained in one complete set $\mathbf{A}$ do not comprise the totality of physical attributes of the system. 
One can form other complete sets, $\mathbf{B},\,\mathbf{C}$ and so on, which are mutually incompatible, and for each choice of noninterfering physical characteristics there is a set of selective measurements referring to systems in the appropriate states'' \cite{schwinger-the_algebra_of_microscopic_measurement}.
From the probabilistic point of view, this means that the groupoid $\Gamma_{A}$ is not enough to encode all the information on the quantum state of the system.
Indeed, in the Hilbert space formulation of quantum mechanics, a (pure) state is represented by a ``ray'', that is, a ``one-dimensional subspace'' of $\mathcal{H}$, associated with non-null vector $|\psi\rangle\in\mathcal{H}$, and when we select a maximal set of compatible observables (resolution of the identity), we are are able to build a probability vector out of $|\psi\rangle$ by setting:
\be
p_{j}(\psi):=\frac{\langle a_{j}|\psi\rangle\langle\psi|a_{j}\rangle}{\langle\psi|\psi\rangle}\,,
\ee
so that $p_{j}(\psi)\geq 0$ and $\sum_{j}\,p_{j}(\psi)=1$.
Clearly, this single probability distribution is not enough to reconstruct the quantum state from which we started.
This is clearly illustrated in the so-called tomographic picture of quantum mechanics \cite{ibort_manko_marmo_simoni_stornaiolo_ventriglia-groupoids_and_the_tomographic_picture_of_quantum_mechanics}, where fair probability distributions, tomograms, are associated with every quantum state, and it is proved that every quantum state may be ``reconstructed'' when a sufficient set, a ``quorum'', of tomograms has been provided \cite{fano-description_of_states_in_quantum_mechanics_by_density_matrix_and_operator_techniques}.
In fact, considering another resolution of the identity $\mathbb{I}=\sum_{k}\,|b_{k}\rangle\langle b_{k}|$, a  new probability vector is associated with $|\psi\rangle$:
\be
p_{k}'(\psi):=\frac{\langle b_{k}|\psi\rangle\langle\psi| b_{k}\rangle}{\langle\psi|\psi\rangle}\,,
\ee
and it is then clear that all the information regarding the possible probability distributions is encoded in the abstract object $\rho=\frac{|\psi\rangle\langle\psi|}{\langle\psi|\psi\rangle}$ which is independent of the  specific resolution of the identity used to define a probability distribution.
We may call this object the ``abstract probability amplitude''.
The replacement of classical fair probability distributions by quantum probability amplitudes is a crucial ingredient to be able to describe such quantum phenomena as interference and entanglement.

We may say that the existence of maximal sets of compatible observables that are mutually incompatible leads to the transition from ``probability distributions'' to ``probability amplitudes''.

Referring again to the bra-ket formalism, by choosing a different observable, say $\mathbf{B}=\sum_{k}\,b_{k}|b_{k}\rangle\langle b_{k}|$,  we obtain another resolution of the identity $\mathbb{I}=\sum_{k}\,|b_{k}\rangle\langle b_{k}|$, and we can build another pair-groupoid $\Gamma_{B}$ whose elements are $M(b_{k},\,b_{l})=|b_{k}\rangle\langle b_{l}|$, and thus another groupoid-algebra associated with $\Gamma_{B}$.
By introducing $U_{AB}:=\sum_{k}\,|a_{k}\rangle\langle b_{k}|$ and noting that $U_{AB}^{\dagger}=U_{BA}=\sum_{k}\,|b_{k}\rangle\langle a_{k}|$, we may build a unitary map that "transforms" the groupoid $\Gamma_{B}$ into the groupoid $\Gamma_{A}$ acting on every $M(b_{j},\,b_{k})\in\Gamma_{B}$ by conjugation with $U_{AB}$, that is, $M(b_{j},\,b_{k})\mapsto U_{AB}\,M(b_{j},\,b_{k})\,U_{AB}^{\dagger}=M(a_{j},\,a_{k})$.
From this, it easily follows that, in the finite-dimensional case, $\Gamma_{A}$ and $\Gamma_{B}$ are isomorphic as groupoids, and thus, their groupoid-algebras are isomorphic as $C^{*}$-algebras.

In Schwinger's work, measurement symbols pertaining to different (incompatible) sets of compatible observables are considered and denoted with the formal symbols $M(a,\,b)$.
These symbols have to encode the fact that ``measurements of properties $\mathbf{B}$, performed on a system in a state $c$ that refers to properties incompatible with $\mathbf{B}$, will yield a statistical distribution of the possible values. Hence, only a determinate fraction of the systems emerging from the first stage will be accepted by the second stage'', and thus the following composition law is introduced:
\be
M(a,\,b)\,M(c,\,d)= <b,\,c>\,M(a,\,d)\,,
\ee
where $<b,\,c>$ is a number characterizing the above-mentioned ``statistical distribution''.
For instance, when the measurement symbols considered pertain to the same maximal set of compatible observables, that is, $b=a$ and $c=a'$, we have $< a,\,a'>=\delta(a,\,a')$.
It is important to introduce the formal symbols $M(a,\,b)$ because, roughly speaking, they can be put into correspondence with composition of Stern-Gerlach devices and they uncover a more profound mathematical structure underlying Schwinger's formalism of atomic measurements which is that of a 2-groupoid.
We will comment thoroughly on this subject in a forthcoming work \cite{ciaglia_ibort_marmo-schwinger}.

In the bra-ket formalism, we may consider the symbols $M(a_{j},\,a_{k})=|a_{j}\rangle\langle a_{k}|$ with $M(b_{l},\,b_{m})=|b_{l}\rangle\langle b_{m}|$ associated with different resolution of the identity, and we may compose them by means of the operator product between $|a_{j}\rangle\langle a_{k}|$ and $|b_{l}\rangle\langle b_{m}|$ obtaining:
\be
M(a_{j},\,a_{k})\,M(b_{l},\,b_{m})=|a_{j}\rangle\langle a_{k}|\,|b_{l}\rangle\langle b_{m}|=\langle a_{k}|b_{l}\rangle\,M(a_{j},\,b_{m})\,.
\ee
It is clear that the quantum mechanical transition probability amplitude $\langle a_{k}|b_{l}\rangle$ plays the role of the statistical coefficient $<a,\,b>$ introduced by Schwinger.
Indeed, remaining in the probabilistic setting, we may review what Schwinger did in terms of abstract ``transition probability amplitudes''.
Indeed, introducing the resolution of the identity $\mathbb{I}=\sum_{j}\,|a_{j}\rangle\langle a_{j}|$, we may write the transition probability $\left|\frac{\langle\psi|\varphi\rangle}{\sqrt{\langle\psi|\psi\rangle}\,\sqrt{\langle\varphi|\varphi\rangle}}\right|^{2}$ between pure quantum states as:
\be
p(\psi,\,\varphi)=\left|\frac{\langle\psi|\varphi\rangle}{\sqrt{\langle\psi|\psi\rangle}\,\sqrt{\langle\varphi|\varphi\rangle}}\right|^{2}=\sum_{j,k}\,\frac{\langle a_{j}|\psi\rangle\langle\varphi|a_{j}\rangle\langle a_{k}|\varphi\rangle\langle\psi|a_{k}\rangle}{\langle\psi|\psi\rangle\,\langle\varphi|\varphi\rangle}\,,
\ee
and similarly we would obtain $p'(\psi,\,\varphi)$ if we change basis.
Again, the abstract mathematical object we would obtain if we get rid of the particular basis would be either
\be
T(\psi,\,\varphi):=\frac{|\psi\rangle\langle\varphi|}{\sqrt{\langle\psi|\psi\rangle}\,\sqrt{\langle\varphi|\varphi\rangle}}\;\;\;\;\mbox{ or } \;\;\;\;
T(\varphi,\,\psi):=\frac{|\varphi\rangle\langle\psi|}{\sqrt{\langle\psi|\psi\rangle}\,\sqrt{\langle\varphi|\varphi\rangle}}\,.
\ee
These two different objects will give rise to the same ``transition probability'' when a basis for the Hilbert space is considered.
We may call these objects ``abstract trasition probability amplitudes''.

When we consider two (incompatible) maximal sets of compatible observables $\mathbf{A}$ and $\mathbf{B}$, we may form the matrix of transition probabilities between the two resolutions of the identity as the matrix whose entries are given by::
\be
p(a_{j},\,b_{k})=\langle a_{j}|b_{k}\rangle\langle b_{k}|a_{j}\rangle \,.
\ee
In Schwinger's terminology \cite{schwinger-the_algebra_of_microscopic_measurement} this matrix is the ``trasformation function'' relating the $\mathbf{A}$-description with the $\mathbf{B}$-description.
It is a stochastic linear map because $p(a_{j},\,b_{k})\geq0$ and $\sum_{j}\,p(a_{j},\,b_{k})=1$.
Furthermore, as $p(a_{j},\,b_{k})=\langle a_{j}|b_{k}\rangle\langle b_{k}|a_{j}\rangle =\langle b_{k}|a_{j}\rangle\langle a_{j}|b_{k}\rangle =p(b_{k},\,a_{j})$ the matrix is symmetric and consequently the associated map is doubly stochastic.

It should be remarked that it is possible to consider  pairs of {\itshape complementary observables}  such that  $p(a_{j},b_{k})$ is independent of the specific $a_{j},\,b_{k}$  and would be equal to $\frac{1}{n}$, if $n$ is the number of different outcomes.
The interpretation is that if  the system is prepared in a state in which the value of $\mathbf{A}$  can be predicted with certainty, then all measurement results are equally probable in a measurement of $\mathbf{B}$.
This feature of quantum mechanics has no true analog in classical physics.

\section{Conclusions}

In conclusion, we have seen how the notion of groupoid naturally emerges when one considers Schwinger's  Symbolism of Atomic Measurements underlying an algebra that turns out to be the a specific representation of the corresponding groupoid-algebra.
This groupoid structure is seen to emerge when we replace transition probabilities with transition amplitudes.
Roughly speaking, this corresponds to passing from $\langle\psi|\varphi\rangle$ to $|\psi\rangle\langle\varphi|$ or $|\varphi\rangle\langle\psi|$.
We have seen how an abstract groupoid structure may be identified in the Hilbert space formalism of quantum mechanics by using an orthonormal discrete resolution of the identity $\mathbb{I}=\sum_{j}\,|a_{j}\rangle\langle a_{j}|$ and defining the elements of the groupoid as $M(a_{j},\,a_{k}):=|a_{j}\rangle\langle a_{k}|$.
By using different resolutions of the identity we may build different groupoids that, in the finite-dimensional case, turn out to be isomorphic.
Once we have the groupoid we are able to build its groupoid-algebra which is naturally an involution algebra that, in the finite-dimensional case, coincides with the full $C^{*}$-algebra of the bounded linear operators on the Hilbert space of the system, thus giving a first insight regarding the connection between the groupoid picture of quantum mechanics presentend here and the $C^{*}$-algebraic picture of quantum mechanics generalizing Heisenberg's picture.
All these considerations will be thoroughly examinated in a forthcoming paper \cite{ciaglia_ibort_marmo-schwinger} where the groupoid picture will be taken as the starting point for a formulation of a quantum theory, and the Hilbert space and $C^{*}$-algebra formalisms will be recovered from the groupoid structure alone.

As Schwinger says \cite{schwinger-the_algebra_of_microscopic_measurement}: ``The entire discussion remains restricted to the realm of quantum statics which, in its lack of explicit reference to time, is concerned either with idealized systems such that all properties are unchanged in time or with measurements performed at a common time''.
Evolution in time, as well as the statistical interpretation of the theory, will be taken up elsewhere and its formalization constitutes the essence of ``Schwinger variational principle''.

In conclusion, while in the Schr\"{o}dinger picture we associate a Hilbert space with every quantum system, a $C^{*}$-algebra in the Hisenberg picture, we could say that we associate a groupoid with every quantum system in the Schwinger picture.

\section*{Acknowledgments}

The authors are happy to thank A. P. Balachandran for the useful comments and remarks he made while reading the manuscript.

The authors acknowledge financial support from the Spanish Ministry of Economy and Competitiveness, through the Severo Ochoa Programme for Centres of Excellence in RD (SEV-2015/0554).
A.I. would like to thank partial support provided by the MINECO research project MTM2017-84098-P and QUITEMAD+, S2013/ICE-2801.   G.M. would like to thank the support provided by the Santander/UC3M Excellence Chair Programme 2016/2017.

\end{document}